\documentclass[12pt]{amsart}
\usepackage{amscd}
\usepackage{amsfonts}
\usepackage{epsf}
\usepackage{epic}
\usepackage{eepic}
\usepackage{latexsym}

\begin{document}

\title[Fields on Noncommutative Torus]
{One-loop Effective Potential for scalar and vector fields  on Higher 
Dimensional Noncommutative Flat Manifolds}  
\author{A.A. Bytsenko}
\address{Departamento de Fisica, Universidade Estadual de Londrina,
Caixa Postal 6001, Londrina-Parana, Brazil\,\, {\em E-mail address:} 
{\rm abyts@uel.br}}
\author{A.E. Gon\c calves}
\address{Departamento de Fisica, Universidade Estadual de Londrina,
Caixa Postal 6001, Londrina-Parana, Brazil\,\, {\em E-mail address:} 
{\rm goncalve@uel.br}}
\author{S. Zerbini}
\address{Department of Physics, University of Trento,
Gruppo Collegato INFN di Trento, Sezione di Padova, Italy\,\, 
{\em E-mail address:} 
{\rm zerbini@science.unitn.it}}

\date{November, 2000}

\thanks{We thanks E. Elizalde, K. Kirsten and L. Vanzo 
for discussions. 
First author partially supported by INFN, Gruppo Collegatodi Trento,  by 
 CNPq grant (Brazil), RFFI 
grant (Russia) No 98-02-18380-a, and by GRACENAS grant (Russia) No 6-18-1997.}

\maketitle

\begin{abstract}

The effective potentials 
for massless scalar and vector quantum field theories on D dimensional 
manifold with  p compact noncommutative
extra dimensions are  evaluated by means of dimensional regularization 
implemented by zeta function tecniques. It is found that, the zeta 
function associated with the  one loop operator may not be regular at 
the origin. 
Thus, the related heat kernel trace has a logarithmic term in the short 
$t$ asymptotic expansion. Consequences of this fact are briefly discussed.

\end{abstract}

Quantum field theories on noncommutative spaces generalize the familiar 
structure of conventional field theories. A motivation for considering these
theories is the appearance they make in theory of strings 
\cite{connes,douglas,seiberg,cheung,chu,schomerus,ardalan},
and the fact that they are  consistent theories  by themselves. It has been 
shown, for example, that 
noncommutative gauge theories describe the low energy excitations of open
strings on $D-$ branes in a background Neveu-Schwarz two-form field $B$
\cite{connes,douglas,seiberg}. 

Recently, the noncommutative perturbative dynamics on $D-$ dimensional 
manifolds has been investigated \cite{seiberg1} and  
  the Kaluza-Klein spectrum for interacting scalars
and vector fields has been calculated in \cite{gomis}; it is consistent with 
formulas found in \cite{seiberg1,kiem,fischler}, where connection between 
calculations
in string theory and field theory on noncommutative torus has been made.
The one loop Casimir energy of scalar and vector fields on noncommutative
space ${\Bbb R}^{1,d}\otimes T^2_{\theta}\,\,({\Bbb R}^{1,d}$ is a flat 
$(d+1)-$ dimensional Minkowski space and $T^2_{\theta}$ is two dimensional
noncommutative torus) has been perturbatively calculated in \cite{nam}.

In this paper we shall calculate by zeta-function methods,  the effective 
action (effective potential times the volume of the manifold) for
massless scalar and vector quantum field theories with compact noncommutative
extra dimensions. 

The actions of the massless interacting scalar and pure $U(1)$ vector fields 
we are considering have respectively the forms

$$
S_{({\rm scalar})}=\int\left(\frac{1}{2}(\partial\,\phi)^2+\frac{\lambda}
{r}\phi\star\phi\star...\star\phi\right)d^Dx
\mbox{,}
\eqno{(1)}
$$
$$
\!\!\!\!\!\!\!\!\!\!\!
S_{(vector)}=-\frac{1}{4}\int F_{MN}\star F^{MN}d^Dx
\mbox{.}
\eqno{(2)}
$$
We are shall work on a  manifold $M={\Bbb R}^{1,d}\otimes T_{\theta}^p$, where
$p=D-d-1$, $p$ even and ${\rm dim}\,M =D $.  $T_{\theta}^p$ is a 
noncommutative 
associative algebra, with elements given by ordinary continous functions on
$T_{\theta}^p$ whose product is given by the Moyal bracket or $(\star) -$
product of functions: 

$$
(F\star G )(x)=\exp\left(\frac{i}{2}\theta^{\mu\nu}
\frac{\partial}{\partial \alpha^\mu}\frac{\partial}{\partial \beta^\nu}
\right)F(x+\alpha)G(x+\beta)|_{\alpha=\beta=0}
\mbox{.}
\eqno{(3)}
$$
 Since we have to deal with a renormalizable 
scalar field theory,  $r$, number of scalar fields in the  $(\star) -$
product,  should  be chosen as an 
integer, given by $r=2D/(D-2)$. As well known, the only  choices are 
$D=4$, $r=4$ and $D=6$,
$r=3$ and $D=3$, $r=6$. With regard to the gauge field, the field strength is

$$
F_{MN}=\partial_MA_N -\partial_NA_M -i{\rm g}(A_M\star A_N -
A_N\star A_M)
\mbox{.}
\eqno{(4)}
$$

We are interested in the one-loop approximation  for theories with 
actions (1) and (2). It has been  shown that the operator ${\mathcal L}$ 
related to 
the one-loop contribution \cite{seiberg1}, has a spectrum given by  
$ {\bf k}^2+ \Omega_{(n,\alpha)}^2$,    
where the Kaluza-Klein spectrum has the form 
(see Refs.\cite{gomis,nam}):

$$
\Omega_{(n,\alpha)}^2 = R^{-2}\varphi({\bf n})(1+\Lambda(\theta) 
 R^{2\alpha}
\varphi({\bf n})^{-\alpha})
\mbox{.}
\eqno{(5)}
$$
Here $\theta$ is the non commutative parameter, and  
$\varphi({\bf n})=\sum_{j}R^{-2} n_j^2$, having chosen,  for the sake of 
simplicity, the 
torus radii equal to $R$. Furthermore, we shall maily interested in the 
scalar field case, then  $\alpha =2$, $\Lambda=\frac{\Lambda_S}{\theta^2} $. 
As illustration, 
we shall also consider  $\alpha =3$,
$\Lambda=\frac{\Lambda_V}{\theta^{4}}$, valid  for vector field, but in 
this case, the 
Kaluza-Klein spectrum is only approximated. The  
coefficients $\Lambda_S$ and $\Lambda_V$ depend on the dimension $D$ 
\cite{seiberg1} and are 
proportional to the coupling constants.  

In this paper, we shall make use of dimensional regularization in the  proper
time formalism implemented by zeta-function techniques 
\cite{elizalde,bytsenko}, namely the one-loop effective action is 
regularized by 
means of ($\mu$ is the renormalization parameter)  
$$
W_{\varepsilon}=\frac{1}{2}
\left({\rm log}\,{\rm det}\,{\mathcal L}/\mu^2\right)_{\varepsilon}=-
\frac{1}{2}\mu^{2\varepsilon}\int_0^\infty dt t^{\varepsilon-1} 
\mbox{Tr} e^{-t{\mathcal L}}
\mbox{,}
\eqno{(6)}
$$
where $\varepsilon$ is a small regularization parameter.
Introducing the zeta function by the usual Mellin transform of the heat-kernel,
one has

$$
W_{\varepsilon}=-\frac{1}{2}
\mu^{2\varepsilon} \Gamma(\varepsilon)\zeta({\varepsilon}|{\mathcal L})
\mbox{.}
\eqno{(7)}
$$
If the zeta function of the one-loop approximation operator is regular at
the origin,  the one-loop divergences and the finite part of the 
effective action can be expressed respectively
by means of  the zeta function and its   
derivative evaluated  at the origin.
Thus, for this reason, in the 
following, we shall  investigate the analytic continuation of 
$\zeta(s|{\mathcal L})$.

For large $\Re\,s$, the spectral zeta function associated with the 
operator 
${\mathcal L}$ acting in manifolds with topology $M$ has 
the form

$$
\zeta(s|{\mathcal L}) =\frac{{\rm Vol}({\Bbb R}^{d+1})}{(2\pi)^{d+1}}
\sum_{{\bf n}\in {\Bbb Z}^p/{\{{\bf 0}\}}}
\int_{{\Bbb R}^{d+1}}({\bf k}^2+ \Omega_{(n,\alpha)}^2)^{-s}d^{d+1}k
\mbox{,}
\eqno{(8)}
$$
where ${\bf k}^2=\sum_{j=0}^dk_j^2$, $\{{\bf 0}\}=(0,0,...,0)$.
Making integration over $d^{d+1}k$ in Eq. (9) with the help of formula

$$
\int_{{\Bbb R}^N}\frac{d^N y}{(y^2+Q)^s}=
\pi^{N/2}Q^{N/2-s}\frac{\Gamma(s-N/2)}{\Gamma(s)},
\,\,\, Q>0,\,\,\,\Re\,s>N/2
\mbox{,}
\eqno{(9)}
$$
where $y^2=\sum_j^N y_j^2$ we obtain:

$$
\zeta(s|{\mathcal L})= {\rm Vol}(\Bbb R^{d+1})
\frac{\pi^{\frac{d+1}{2}}}{(2\pi)^{d+1}}
\frac{\Gamma\left(s-\frac{d+1}{2}\right)}
{\Gamma(s)R^{d+1-2s}}
\sum_{{\bf n}\in {\Bbb Z}^p/{\{{\bf 0}\}}}
\varphi({\bf n})^{\frac{d+1}{2}-s}
$$
$$
\!\!\!\!\!\!\!\!\!\!\!\!\!\!\!\!\!\!\!\!\!\!\!
\!\!\!\!\!\!\!\!\!\!\!\!\!\!\!\!\!\!\!\!\!\!\!
\times
\left[
1+ \Lambda \theta^{2-2\alpha}R^{2\alpha}
\varphi({\bf n})^{-\alpha}\right]^{\frac{d+1}{2}-s}
\mbox{.}
\eqno{(10)}
$$

The analytic continuation can be achieved by using the binomial expansion
(\cite{elizalde,bytsenko}). As a result, for 
$ C=\Lambda(\theta)R^{2\alpha}<1 $, one has
$$
\zeta(s|{\mathcal L}) = A(s;d)
\sum_{\ell=0}^{\infty}
\frac{(-C)^{\ell}\Gamma\left(s-\frac{d+1}{2}+\ell \right)}
{\ell!\Gamma\left( s\right)}
Z_p\left| \begin{array}{ll}
{\bf 0}\\
{\bf 0}\\
\end{array} \right|\left(2s+p+2\ell \alpha-D\right)
\mbox{,}
\eqno{(11)}
$$
where we have introduced the quantity
$$
A(s;d)=
{\rm Vol}({\Bbb R}^{d+1})\frac{\pi^{\frac{d+1}{2}}}{(2\pi)^{d+1}}
R^{2s-d-1}
\mbox{.}
\eqno{(12)}
$$
In Eq. (11),  $Z_p\left|_{\bf h}^{\bf g}\right|(s,\varphi)$ is the 
$p$-dimensional 
Epstein zeta function associated with the quadratic form 
$\varphi [a({\bf n}+{\bf g})]=\sum_ja_j(n_j+{\rm g}_j)^2.$ For $\Re\,s>p$\,\,
$Z_p\left|_{\bf h}^{\bf g}\right|(s,\varphi)$ is given by the formula

$$
Z_p\left| \begin{array}{ll}
{\rm g}_1\,...\,{\rm g}_p \\
h_1\,...\,h_p\\
\end{array} \right|(s,\varphi)=\sum_{{\bf n}\in {\bf Z}^p}{}'
\left(\varphi[a({\bf n}+{\bf g})]\right)^{-s/2}
$$
$$
\times\exp\left[2\pi i({\bf n},{\bf h})\right]
\mbox{,}
\eqno{(13)}
$$
where ${\rm g}_j$ and  ${\rm h}_j$ are some real numbers \cite{bateman}, the
prime means omitting the term with $(n_1, n_2, ..., n_p)=
({\rm g}_1, {\rm g}_2, ..., {\rm g}_p)$ if all the ${\rm g}_j$ are integers.
The functional equation for 
$Z_p\left|_{\bf h}^{\bf g}\right|(s,\varphi)$ reads,

$$
Z_p\left| \begin{array}{ll}
{\bf g}\\
{\bf h}\\
\end{array} \right|(s,\varphi)=
({\rm det}\,a)^{-1/2}\pi^{\frac{1}{2}(2s-p)}\frac{\Gamma(\frac{p-s}{2})}
{\Gamma(\frac{s}{2})}\exp[-2\pi i({\bf g},{\bf h})]
$$
$$
\times
Z_p\left| \begin{array}{ll}
{\bf\,\,\,\, h}\\
-{\bf g}\\
\end{array} \right|(p-s,\varphi^*)
\mbox{,}
\eqno{(14)}
$$
where $\varphi^*[a({\bf n}+ {\bf g})]=\sum_j a_j^{-1}(n_j+{\rm g}_j)^2$.

Formulae (11) and (14) give the analytic continuation of the zeta
function. 

Since the Epstein zeta-function has an analytic estension with only a simple 
pole at $s=p$, it follows easily that the regular values of the zeta function 
at the origin are present when $D \neq 2\alpha k$ with  $k\in {\Bbb Z}_{+}$, 
namely 
$$
\lim_{s\rightarrow 0 \atop {d=2k}}\zeta(s|{\mathcal L})= 0
\mbox{,}
\eqno{(15)}
$$
$$
\lim_{s\rightarrow 0 \atop {d=2k-1}}
\zeta(s| {\mathcal L})=
A(0;2k-1)\sum_{\ell=1}^{k+1}\frac{(-C)^\ell (-1)^{(k+1-\ell)}}
{\ell!(k+1-\ell)!}
Z_p\left| \begin{array}{ll}
{\bf 0}\\
{\bf 0}\\
\end{array} \right|\left(p+2\ell \alpha-D\right)
\mbox{,}
\eqno{(16)}
$$
where

$$
A(0;2k-1)=\frac{{\rm Vol}({\Bbb R}^{2k})}{(4\pi)^k R^{2k}}
\mbox{.}
\eqno{(17)}
$$
It should be noted that in presence of the  noncommutative torus, the values 
of the zeta 
function in the point 
$s=0$ differ  from the ones  in the case of torus, where
$\zeta(0|{\mathcal L})=0$ for$D$ odd (see for example 
\cite{bytsenko1}).

For these cases, the finite part of effective action is given by the sum of 
the zeta function
and its derivative evaluated at the origin. Thus, the  one-loop  effective 
action, after renormalization,   has the form

$$
W=-\frac{1}{2}\left(\frac{d}{ds}
\zeta(s|{\mathcal L})|_{s=0}
+ \zeta(0|{\mathcal L}){\rm log}(\mu^2)\right)
\mbox{,}
$$
where, for $d=2k$, one has
$$
\zeta'(0|{\mathcal L}) = A(0;2k)
\sum_{\ell=0}^{\infty}
\frac{(-C)^{\ell}\Gamma\left(-\frac{d+1}{2}+\ell \right)}
{\ell!}
Z_p\left| \begin{array}{ll}
{\bf 0}\\
{\bf 0}\\
\end{array} \right|\left(p+2\ell \alpha-D\right)
\mbox{,}
\eqno{(18)}
$$
while for $d=2k-1$,
$$
\zeta'(0|{\mathcal L}) = 2 A(0;2k-1)
\sum_{\ell=0}^{k+1}
\frac{(-C)^{\ell}(-1)^{k+1+\ell}}
{\ell!(k+1-l)!}
Z_p'\left| \begin{array}{ll}
{\bf 0}\\
{\bf 0}\\
\end{array} \right|\left(p+2\ell \alpha-D\right)
$$
$$
+\ln R^2 \zeta(0|{\mathcal L})+
A(0;2k-1)
\sum_{\ell=1}^{k+1}d_\ell
\frac{(-C)^{\ell}(-1)^{k+1+\ell}}
{\ell!(k+1-l)!}
Z_p\left| \begin{array}{ll}
{\bf 0}\\
{\bf 0}\\
\end{array} \right|\left(p+2\ell \alpha-D\right)
$$
$$
+ A(0;2k-1)
\sum_{\ell=k+2}^{\infty}
\frac{(-C)^{\ell}\Gamma(\ell-k-1)}
{\ell!(k+1-l)!}
Z_p\left| \begin{array}{ll}
{\bf 0}\\
{\bf 0}\\
\end{array} \right|\left(p+2\ell \alpha-D\right)
\mbox{,}
\eqno{(19)}
$$
with
$$
d_\ell=\sum_{j=1}^{k+1-\ell}\frac{1}{j}
\mbox{,}
\eqno{(20)}
$$
and $\zeta(0|{\mathcal L})$ given by Eq. (16).
Also for $d$ even, and  $D=2\alpha k$, the zeta at the origin is finite and 
different from zero. In fact, we have
$$
\zeta(0|{\mathcal L}) = -2 A(0;2k)\sqrt \pi \frac{\pi^{p/2}}
{\Gamma(p/2)}(-C)^{k}
\mbox{.}
\eqno{(21)}
$$

However, for $d$ odd and $D=2\alpha k$\,, $\zeta(s|{\mathcal L})$
may have a pole at the origin. This is quite unsual. 
To our knowledge, this has also 
been noted investigating  quantum field theory on generalized cone 
\cite{kirsten,cognola} and in ultrastatic 4-dimensional space-times 
with non-compact, but finite volume, 3-dimensional  hyperbolic spatial 
section \cite{bytsenko2}. 

For scalar fields, $\alpha=2$. Thus there exist solutions of the 
constraint $D=d+1+p$, with $p$ even, given by $D=4k$, $d=4k-p-1>0$ and 
$p=2q, q\in {\Bbb Z}_+$.
In these cases, the zeta function has  a pole at the origin. 

As an example, let us consider in detail the case $D=4$, $p=2$ $k=1$, $d=1$.
One has
$$
\zeta(s|{\mathcal L}) = A(s;1)
\sum_{\ell=0, \ell \neq 1}^{\infty}
\frac{(-C)^{\ell}\Gamma\left(s-1+\ell \right)}
{\ell!\Gamma\left( s\right)}
Z_2\left| \begin{array}{ll}
{\bf 0}\\
{\bf 0}\\
\end{array} \right|\left(2s+4 \ell-2\right)+
$$
$$
+A(s;1)(- C) Z_2\left| \begin{array}{ll}
{\bf 0}\\
{\bf 0}\\
\end{array} \right|\left(2s+2\right)
\mbox{.}
\eqno{(22)}
$$
Here, one can show the simple pole for $s=0$, due to the last term in the 
above equation.
The presence of this pole, depends on the constant $C$, namely on the 
coupling constant and on the non commutative parameter $\theta$ and does not 
depend on the compactness of the noncommuting coordinates. 

In fact, if we consider a massless scalar 
field defined on  ${\Bbb R}^{1,d}\otimes {\Bbb R}^p_{\theta}$, namely on 
${\Bbb R}^D$, with $p$ noncommuting coordinates, a direct calculation gives 
the {\it exact} formula for the analytic continuation of zeta function:
$$  
\zeta(s|{\mathcal L})=\frac{{\rm Vol}({\Bbb R}^{D})}{2(4\pi)^{D/2}
\Gamma(p/2)}
\frac{\Lambda_S^{\frac{D-2s}{4}}}{\theta^{\frac{D-2s}{2}}\Gamma(s)}
\Gamma(\frac{2s+p-d-1}{4})
\Gamma(\frac{2s-D}{4})
\mbox{.}
\eqno{(23)}
$$
Eq. (23) is in agreement with the perturbative one obtained in the 
case of the compact noncommutative p-dimensional torus, since for $d=4j+p-1$ 
and
$D=4j+2p$, $j=0,1,2,..$ there is a pole at the origin.  It should be noted that
there is no pole at the origin for the zeta function if {\it all} 
the coordinates of ${\Bbb R}^{D}$ are noncommuting, namely if  $d+1=0$.

Again, for $D=4$, $p=2$, $d=1$, one has

$$  
\zeta(s|{\mathcal L})=\frac{{\rm Vol}({\Bbb R}^{4})}{2(4\pi)^{2}}
\frac{\Lambda_S^{1-s/2}}{\theta^{2-s}}
\frac{\Gamma(\frac{s}{2})}{\Gamma(s)}
\Gamma(\frac{s}{2}-1)
\mbox{.}
\eqno{(24)}
$$

Let us analyze the consequences of the presence of this pole with regard to 
the structure of the one-loop divergences. 
For small $\varepsilon$, one has
$$
W_{\varepsilon}=-\frac{1}{2}
\mu^{2\varepsilon}\Gamma(\varepsilon)\zeta({\varepsilon}|{\mathcal L})
=\frac{B_0}{\varepsilon^2}+\frac{B_1}{\varepsilon}+B_2+{\mathcal O}
(\varepsilon)
\mbox{,}
\eqno{(25)}
$$  
where the coefficients $B_j$ can be computed explicity. In fact, one has:
$$
B_0=-\frac{1}{2}{\rm Res}
\,\left(\zeta(\varepsilon|{\mathcal L})|_{\varepsilon =0}\right)\,
\mbox{,}
\eqno{(26)}
$$
$$
B_1=-\frac{1}{2}{\rm lim}_{\varepsilon\rightarrow 0}\frac{d}{d\varepsilon}
\left(\mu^{2\varepsilon}\varepsilon \Gamma(1+\varepsilon)
\zeta(\varepsilon|{\mathcal L})\right)
\mbox{,}
\eqno{(27)}
$$
$$
B_2=-\frac{1}{4}{\rm lim}_{\varepsilon\rightarrow 0}\frac{d^2}{d^2\varepsilon}
\left(\mu^{2\varepsilon} \varepsilon \Gamma(1+\varepsilon)
\zeta(\varepsilon|{\mathcal L})\right)
\mbox{.}
\eqno{(28)}
$$
As an example, for the explicit and  non compact 
4-dimensional case  we have 
considered, Eq. (22) gives
$$
B_0=\frac{{\rm Vol}({\Bbb R}^{4})}{(4\pi)^{2}}
\frac{\Lambda_S}{\theta^{2}}
\mbox{,}
\eqno{(29)}
$$
$$
B_1=\frac{{\rm Vol}({\Bbb R}^{4})}{(4\pi)^{2}}
\frac{\Lambda_S}{\theta^{2}}\left( \ln \frac{\mu^2\theta}{\sqrt \Lambda_s}
+\frac{1}{2}-\gamma \right)
\mbox{,}
\eqno{(30)}
$$
$$
B_2=\frac{{\rm Vol}({\Bbb R}^{4})}{4(4\pi)^{2}}
\frac{\Lambda_S}{\theta^{2}}\left( (\ln \frac{\mu^2\theta}{\sqrt \Lambda_s}
+\frac{1}{2}-\gamma)^2 +\Psi'(1)+\frac{1}{2}\right)
\mbox{,}
\eqno{(31)}
$$
where  $\Psi(z)$ is the logarihmic derivative of the Gamma 
function and $\gamma=-\Psi(1)$ is the Euler-Mascheroni constant.
As a consequence, the one-loop renormalization appears to be  problematic,
due to the presence of $\mu$ in $B_1$. However, with regard to  issue of the 
renormalization of scalar fields on 
noncommutative space, see \cite{roiban}.

Let us conclude with some remarks. In this letter, we have investigate the 
one-loop approximation for scalar and gauge fields defined on the Euclidean  
 manifold $M={\Bbb R}^{1,d}\otimes T_{\theta}^p$, making use of dimensional 
regularization and zeta function tecniques. The non compact case has also 
been considered. In order to find the analytic 
continuation of the zeta function related to the one-loop operator, in the 
compact case, we have 
used the binomial theorem and our results are valid for $C <1$. This condition
is not so restrictive, because leads to small compactification radius as soon 
as one is dealing with small non commutative parameter $\theta$. 

 We have   found that when 
$d$ is odd and $D=4,8,12,..$ for scalar fields and 
$D=6,12,18,..$ for vector fields, the zeta function related
to  the one loop operator contains 
a pole at the origin. Similar results have been obtained in the non compact 
case. This fact is  mainly associated with  the  
presence of pairs of noncommutative coordinates, compact or non compact. 

Furthermore, the presence of the pole implies that the Casimir or
vacuum energy is singular, when the analytic regularization parameter is 
removed.  This feature is in common with 
the  quantization of  fields defined on  
manifolds  with higher dimensional conical singularities and in 4-dimensional 
ultrastatic space-time with non-compact, but finite volume,  hyperbolic 
spatial section. 
With regard to last issue and in the presence of compact noncommuting 
coordinates , 
the role of the self-consistency determination of the internal radius 
\cite{candelas} is 
also  present in the models we have considered. When the pole in the zeta 
function is absent, one may try to use the presence of the vacuum energy 
as stabilization mechanism for the compactification radius 
\cite{nam,huang}. 
However, when the pole is present, first one has to do a careful 
and  non trivial renormalization analysis.  

As far as this issue is concerned, another consequence of this 
pole is the appearance of logarithmic term $\ln t$
in the short $t$ heat kernel expansion of $\mbox{Tr} e^{-t {\mathcal L}}$, 
rendering ${\mathcal L}$ a pseudo-differential operator 
(see \cite{cognola2,gilkey}) and 
thus modyfing the 
structure of the ultraviolet divergences of the one-loop effective action.
We have computed these divergences and the finite part of the 
effective action. The knowledge of these divergences, which seem  more severe 
than the ordinary cases, could be  helpful in the discussion concerning the 
one-loop renormalizability of 
the models, when the space-time is curved or has a non-trivial topology and
in the issue of the self-consistent determination of the compactification 
radius \cite{candelas}.

\end{document}